\documentclass[twocolumn,elsart1p.cls,floatfix]{revtex4}
\usepackage{color}
\usepackage{epsfig}

\begin{document}

\title{Melting of alloys along the  inter-phase boundaries in eutectic and peritectic
systems}
\author{Efim A. Brener and D. E. Temkin}
\affiliation{Institut f\"ur Festk\"orperforschung, Forschungszentrum J\"ulich,
D-52425 J\"ulich, Germany} 

\begin{abstract}

We discuss a  simple model of the melting kinetics 
along the solid-solid interface  in eutectic and peritectic systems. 
The process is controlled by the diffusion inside the liquid phase and the existence 
of a triple junction is crucial  for the velocity selection problem.
Using the lubrication approximation for the diffusion field in the liquid phase 
we obtain scaling results for the  steady-state velocity of 
the moving pattern depending on the overheating above the equilibrium temperature and 
on the material parameters of the system, 
including the dependences on the angles at the triple junction. 

\end{abstract}

\author{Keywords: melting of alloys, dendritic growth, diffusion, 
kinetics self-organization and patterning}

\maketitle
\section{Introduction}
 The systematic  investigation of  melting kinetics in alloys, and  particular 
in eutectic and peritectic systems, is much less developed than the  
investigation of solidification (for a recent review on solidification see, for example 
\cite{MKB02} and references therein).  Microstructures, 
being at the center of materials science and engineering, are formed during 
the solidification process and, in this sense, the melting process is less attractive 
for practical applications. 

However,  the interfacial pattern selection problem  during the melting 
process might be  very interesting. For example, in our previous publications 
\cite {temkin2005, brener2005} we discussed a free boundary problem for two 
moving interfaces that strongly interact via the diffusion field in the liquid layer 
between them.  This problem arises in the context of liquid film migration during the 
partial melting of solid alloys \cite{musch1} and  could also be relevant to the  
sintering process in the presence of the liquid phase \cite{yoon1}. 
For the melting of one-phase alloys  to proceed in this way,   
 the local equilibrium concentrations have to be different for  the two interfaces 
 providing the driving force for the process. 
 It is by now well accepted  (see, for example, \cite{yoon2,4}) 
that the difference of the equilibrium states at the melting and solidification fronts 
is due to the coherency strain energy which is   
important only at the melting front because of the sharp concentration profile  
ahead the moving melting front. 
 
The other source of  elastic deformations during the melting process, even in 
pure materials, is the density difference between the solid and the liquid phase. 
If the melt inclusion 
is entirely inside  the solid matrix,  inhomogeneous elastic deformations 
inevitably arise. The peculiar behavior of the melting kinetics in such systems was  
discussed in \cite{bm,bmi}.
  
\begin{figure}
\begin{center}
\epsfig{file=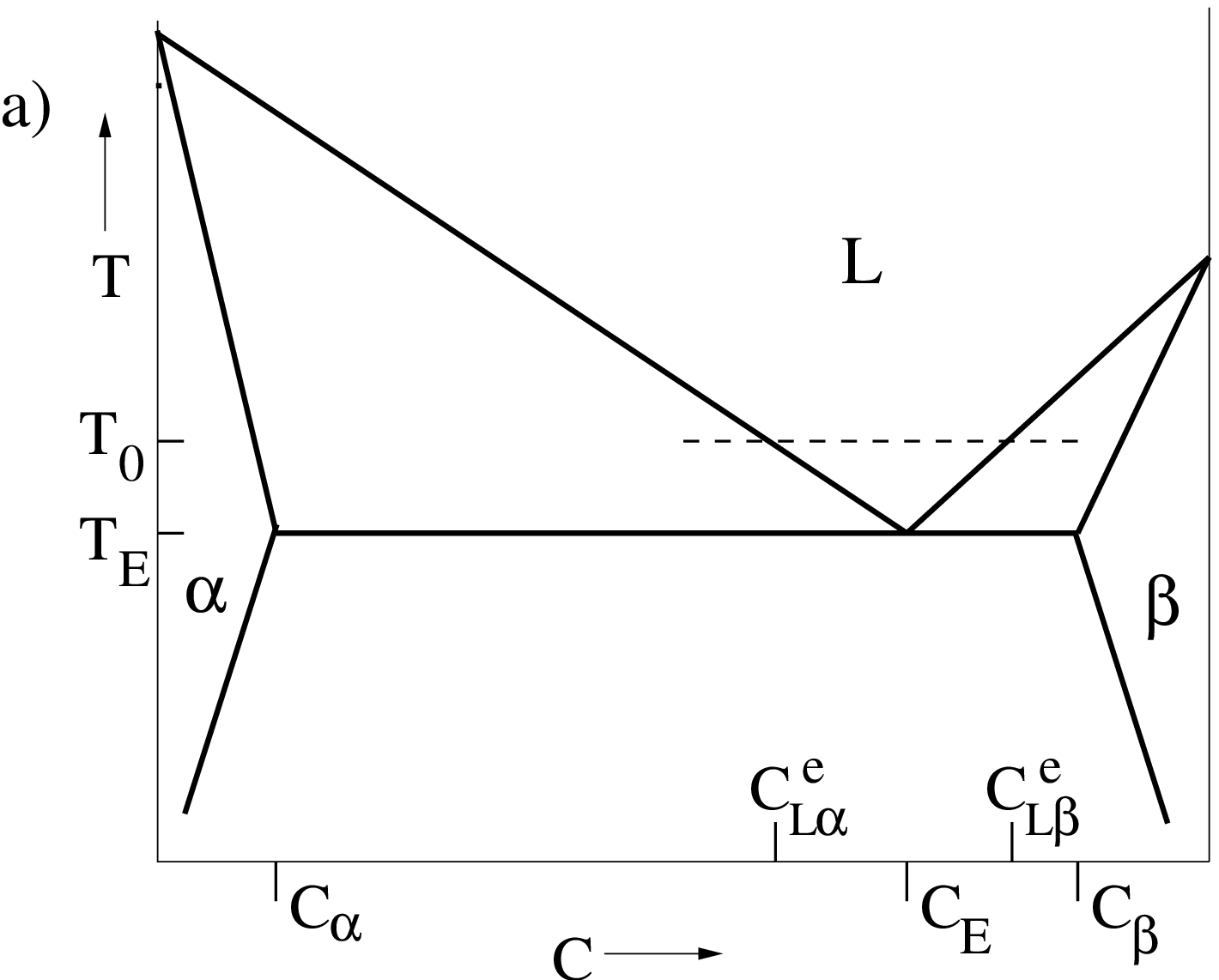, width=5.5cm}
\epsfig{file=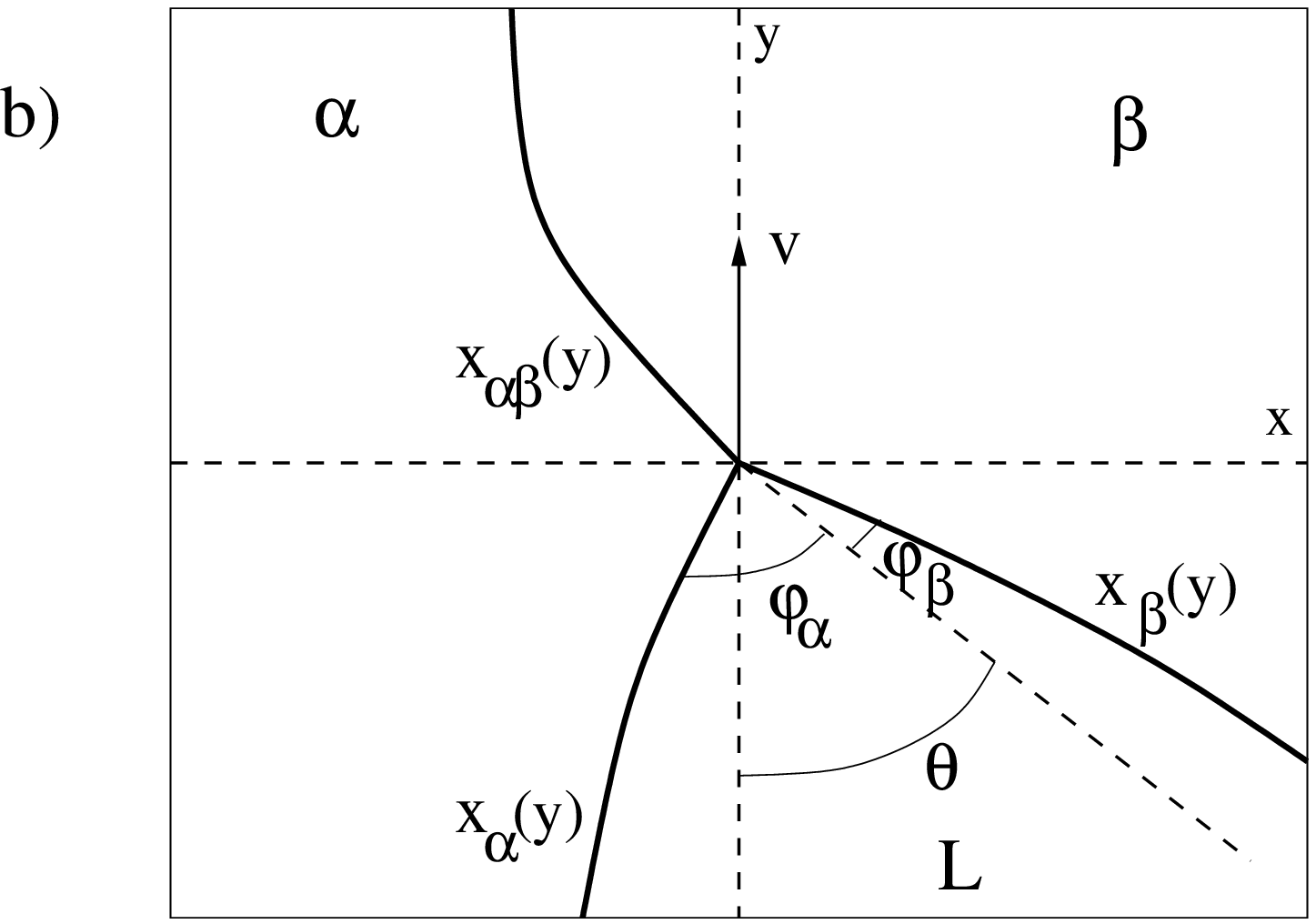, width=5.5cm}
\caption{Schematic presentation of the phase diagram a) and 
 configuration of different interfaces near the triple junction b) in eutectic systems.
$x_{\alpha\beta}(y)$ is the interface between two solid phases $\alpha$ and $\beta$; 
the interfaces between the liquid phase $L$ and  the solid phases are denoted by 
$x_{\alpha}(y)$ and $x_{\beta}(y)$. In the steady state regime this configuration moves 
along the $y$ axis with a constant velocity $v$.  
The origin of the coordinate system is located at 
the triple junction.}
\end{center}
\end{figure}

The main purpose of this paper is to describe the problem of contact melting in  
eutectic and peritectic systems along the boundary between two solid phases (see Fig. 1 
and Fig. 2). The local  concentrations at the $L/\alpha$ and $L/\beta$ interfaces 
in such systems are  different because of the chemical difference between the  
$\alpha$ and $\beta$ phases, 
and weak coherency strain effects are not so important here. If we also assume that the 
liquid phase extends up to the sample surfaces, the mentioned elastic deformations due  
 the density difference are not important either,  
because a weak hydrodynamic flow inside  the liquid phase  compensates  
 the density difference.
 
We concentrate here on the velocity selection problem during the melting 
along the solid-solid interface. The presence of the triple 
junction (see Fig.1b and Fig. 2b) plays a crucial role in this process. 
In the classical problem of 
dendritic growth, where the triple junction is not present, the velocity selection 
is controlled by tiny singular effects of the anisotropy of the surface energy 
(for review see \cite{kessler88,brener91}). The triple junction produces  a very strong 
perturbation of the liquid-solid interfaces and controls the velocity selection. 
The other important difference compared to the classical dendritic problem  
is that the kinetics of the contact melting in eutectic and peritectic systems 
 is controlled by the diffusion inside the needle-like liquid phase and not in the outer
phase.

All features of the contact melting process  mentioned so far are common for both 
systems. There is, however, important difference between  eutectic and  peritectic 
systems. In the eutectic system (Fig. 1)  both interfaces $L/\alpha$ and $L/\beta$ 
are  melting fronts while in the peritectic system (Fig.2) the $L/\alpha$ interface is 
a solidification front if the temperature $T_0$ is above the peritectic temperature 
$T_P$. In  other words, during the melting of peritectic  systems the low-temperature 
$\beta$-phase melts while the high-temperature $\alpha$-phase solidifies. It means 
that, additionally  
to the formation of the liquid phase, the polymorphic transition $\beta\rightarrow\alpha$
occurs. In this context the contact melting in the peritectic system is similar 
to the process of liquid film migration, mentioned above, 
where  the liquid film  is also located between  melting and solidification fronts.

Finally, we  discuss in this paper the evolution 
of the solid-solid interface due to surface diffusion.
This process inevitably arises due to the deviation  of the interface from a flat
configuration in the vicinity of a triple junction. We show, however, that  surface 
diffusion  does not play a controlling role in the melting kinetics, 
and only allows for the necessarily adjustment of the solid-solid interface.

\section{Isothermal melting in eutectic system}

We consider  the two-dimensional problem of the simultaneous melting of two eutectic 
phases $\alpha$ and $\beta$ along the boundary between them.
The phase diagram and the configuration of different interfaces near the triple junction 
 are schematically presented in Fig.1. The compositions of the solid phases are close 
to their equilibrium values $C_{\alpha}$ and $C_{\beta}$ at the eutectic temperature.  
We assume that  diffusion in 
the solid phases is very slow. The temperature of the sample,  $T_0$, is  
slightly above the eutectic temperature $T_E$ (for the notations see Fig.1).
At small overheating the concentration field in the liquid, $C(x,y)$ obeys  
the Laplace equation
\begin{equation} \label{laplace}
\Delta C=0.
\end{equation}
 We assume  local equilibrium at the liquid-solid interfaces. 
Then the concentrations at the $L/\alpha$ and $L/\beta$ 
interfaces are:
\begin{eqnarray}\label{equil}
C_{L\alpha}(y)= C_{L\alpha}^e+(C_{\beta}-C_{\alpha})d_{\alpha}x_{\alpha}''(y),\nonumber\\
C_{L\beta}(y)= C_{L\beta}^e+(C_{\beta}-C_{\alpha})d_{\beta}x_{\beta}''(y).
\end{eqnarray}
Here the notations of the different concentrations are clear from the 
phase diagram (Fig. 1a); the capillary lengths are 
$$d_{\alpha(\beta)}=\frac{\sigma_{\alpha(\beta)}\Omega}{|C_E-C_{\alpha(\beta)}|
(C_{\beta}-C_{\alpha})f_L''(C_E)},$$
where $\sigma_{\alpha(\beta)}$ is the surface energy of the $L/\alpha(\beta)$ interface,
$\Omega$ is atomic volume which is assumed to be the same in all three phases,
$f_L(C)$ is the free energy  of the liquid phase per atom and $f_L''(C)$ is its second 
derivative with respect to $C$.  
  The mass balance conditions at these interfaces read:
\begin{eqnarray}\label{mass}
D\partial C/\partial x = v(C_E-C_{\alpha})x_{\alpha}'(y),\nonumber\\
D\partial C/\partial x = -v(C_{\beta}-C_E)x_{\beta}'(y),
\end{eqnarray}
where $D$ is the diffusion coefficient in the liquid phase.
 
Eqs. (\ref{equil})-(\ref{mass}) are written for the case of small angles 
$\theta$, $\varphi_{\alpha}$ and $\varphi_{\beta}$ at the triple junction. 
This leads to the small angles along the whole interface, $x_{\alpha}^{'}\ll1$
and $x_{\beta}^{'}\ll1$. In this case we can use the so-called 
``lubrication'' approximation for the solution of the Laplace equation. 
In this approximation one neglects the derivatives 
with respect to the ``slow'' $y$ variable as compared with the derivatives with respect 
to the ``fast'' $x$ variable:

\begin{equation}\label{concentration}
C(x,y)=C_0(y)+B(y)x.
\end{equation}

 The slow variable functions $C_0(y)$ and $B(y)$ can be found from 
the boundary condition, Eq. (\ref{equil}):
\begin{eqnarray}\label{constant}
 C_0(y)=\frac{C_{L\alpha}(y)x_{\beta}(y)- C_{L\beta}(y)x_{\alpha}(y)}
{x_{\beta}(y)-x_{\alpha}(y)},\nonumber \\
 B(y)=\frac{C_{L\beta}(y)- C_{L\alpha}(y)}
{x_{\beta}(y)-x_{\alpha}(y)}.
\end{eqnarray}
From Eqs. (\ref{mass}) and (\ref{concentration}) we find the following relations,
\begin{equation}\label{prof}
x_{\alpha}^{'}=\frac{D}{v}\frac{B(y)}{(C_E-C_{\alpha})}, \,\,\,
x_{\beta}^{'}=-\frac{D}{v}\frac{B(y)}{(C_{\beta}-C_E)},
\end{equation}
which together with Eqs. (\ref{constant}) and (\ref{equil}) form a closed system of 
 two second order differential equations for the front profiles $x_{\alpha}(y)$ and 
$x_{\beta}(y)$. They are subject to the boundary conditions at the triple junction:
\begin{eqnarray}\label{bound1}
x_{\alpha}(0)=x_{\beta}(0)=0,
\end{eqnarray}
\begin{eqnarray}\label{bound2}
x_{\alpha}^{'}(0)=\varphi_{\alpha}-\theta\equiv\theta_{\alpha}; \,\,\,
 x_{\beta}^{'}(0)=-(\varphi_{\beta}+\theta)\equiv -\theta_{\beta}
\end{eqnarray}
(see Fig. 1b). It follows from Eqs. (\ref{prof}) and (\ref{bound1}) that 
\begin{equation}\label{match}
x_{\alpha}(y)=-gx_{\beta}(y),
\end{equation} 
where $g=(C_{\beta}-C_E)/(C_E-C_{\alpha})$. 
 We also find from 
Eq. (\ref{bound2}) that
\begin{equation}\label{angle}
\theta_{\alpha}=\frac{g(\varphi_{\alpha}+\varphi_{\beta})}{1+g},\,\,\,
\theta_{\beta}=\frac{\varphi_{\alpha}+\varphi_{\beta}}{1+g},\,\,\,
\theta=\frac{\varphi_{\alpha}-g\varphi_{\beta}}{1+g}.\,\,\,
\end{equation}
The relation (\ref{match}) allows to eliminate the profile $x_{\alpha}$ and to write 
 the closed equation for the profile $x_{\beta}$. Moreover, one can integrate this 
equation once with respect to $y$ and finally obtains a nonlinear 
first order differential equation for the profile $x_{\beta}$:
\begin{equation}\label{first}
-(d_{\beta}+gd_{\alpha})x_{\beta}^{'}=(d_{\beta}+gd_{\alpha})\theta_{\beta}
+\Delta y+
\frac{vg}{2D}x_{\beta}^{2}.
\end{equation}
Here $\Delta=(C_{L\beta}^e-C_{L\alpha}^e)/(C_{\beta}-C_{\alpha})$ is the dimensionless
overheating above the eutectic temperature.

Let us introduce the dimensionless coordinates:
\begin{equation}\label{rescale}
X_{\beta}=\frac{x_{\beta}\Delta}{\theta_{\beta}^{2}(d_{\beta}+gd_{\alpha})},\,\,\,
Y=\frac{y\Delta}{\theta_{\beta}(d_{\beta}+gd_{\alpha})}.
\end{equation}
Then Eq. (\ref{first}) takes the form
\begin{equation}\label{basic}
-X_{\beta}^{'}=1+Y+\nu X_{\beta}^{2}
\end{equation}
with the dimensionless parameter $\nu$: 
\begin{equation}\label{parameter}
\nu=v\frac{g(d_{\beta}+gd_{\alpha})\theta_{\beta}^{3}}{2D\Delta^2}.
\end{equation}  
The solution of this equation starts at the origin with $X_{\beta}^{'}(0)=-1$ and 
should have  parabolic asymptotics $X_{\beta}=\sqrt{|Y|/\nu}$
for large values of $|Y|$. It turns out that 
such a smooth solution exists only if the parameter $\nu=\nu^{\star}\approx 1.06$.   
In other words, this is a nonlinear eigenvalue problem 
which leads to the selection of the velocity $v$:
\begin{equation}\label{velocity}
 v=\nu^{\star}\frac{2D\Delta^2}{g(d_{\beta}+gd_{\alpha})\theta_{\beta}^{3}}
\end{equation}
We see that the posed problem has a relatively simple, essentially analytical solution.

\section{Isothermal melting in peritectic system}

\begin{figure}
\begin{center}
\epsfig{file=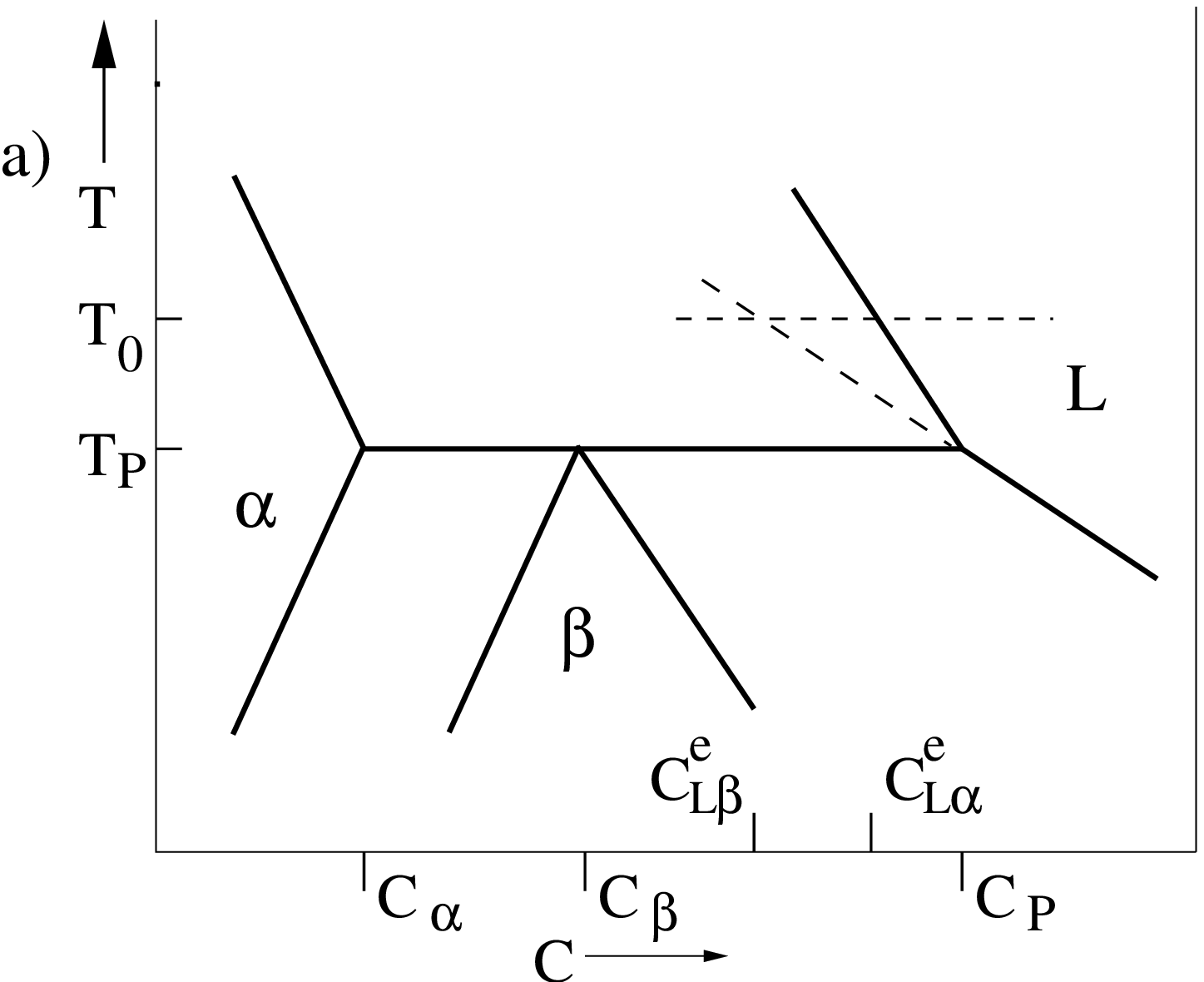, width=5.5cm}
\epsfig{file=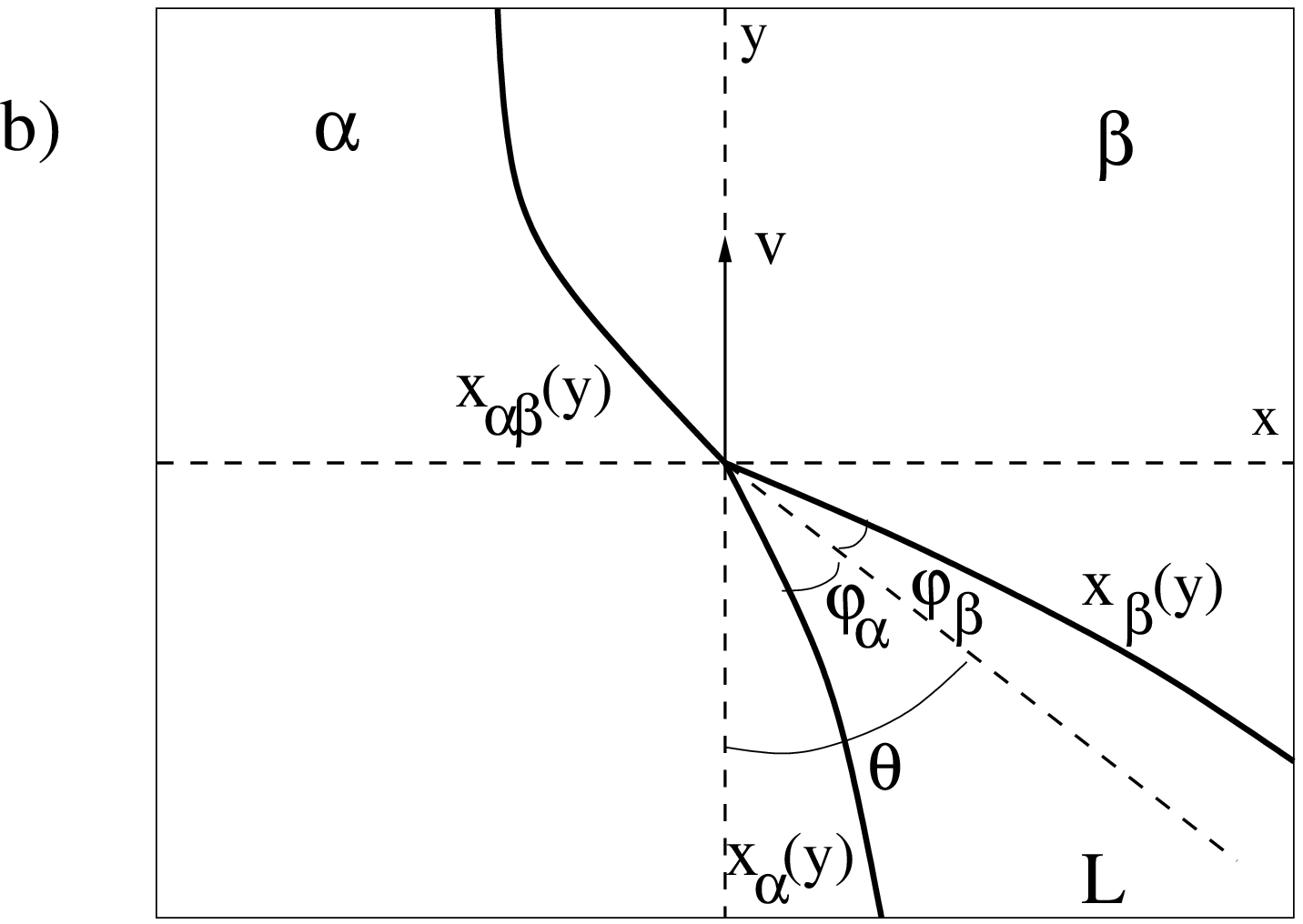, width=5.5cm}
\caption{Schematic presentation of the phase diagram a) and configuration 
of different interfaces near the triple junction b) in peritectic systems.
$x_{\alpha\beta}(y)$ is the interface between two solid phases $\alpha$ and $\beta$; 
the interfaces between the liquid phase $L$ and  the solid phases are denoted by  
$x_{\alpha}(y)$ and $x_{\beta}(y)$. In the steady state regime this configuration moves 
along the $y$ axis with a constant velocity $v$.  
The origin of the coordinate system is located at 
the triple junction.}
\end{center}
\end{figure}

The phase diagram and the configuration of different interfaces near the triple junction 
 are schematically presented in Fig.2.  The temperature of the sample is $T_0$ and 
slightly above the peritectic temperature $T_P$ (for the notations see Fig.2a).
Overheating is assumed to be small and the compositions of the solid phases are 
close to the equilibrium values $C_{\alpha}$ and $C_{\beta}$ 
at the peritectic temperature. Eqs. (\ref{laplace})-(\ref{mass}) are valid also 
for the peritectic system where $C_E$ should be replace by $C_P$ and in the second 
equation in (\ref{equil}) one should change the sign in front of the capillary 
term,  
$$
C_{L\beta}(y)= C_{L\beta}^e-(C_{\beta}-C_{\alpha})d_{\beta}x_{\beta}''(y).
$$
This reflects the fact that the equilibrium between the $\beta$-phase and the 
liquid phase
corresponds to the negative slope of the liquidus line for the peritectic phase diagram 
and to the positive slope for the eutectic diagram.
As a result, Eqs. (\ref{match})-(\ref{angle}) remain the same and Eq. (\ref{first}) is 
replaced by
\begin{equation}\label{firstp}
 -(d_{\beta}-gd_{\alpha})x_{\beta}^{'}=(d_{\beta}-gd_{\alpha})\theta_{\beta}
-\Delta y-
\frac{vg}{2D}x_{\beta}^{2}.
\end{equation}
 
We note that the parameters $g$ and $\Delta$ for the peritectic system (Fig. 2) are
negative if we define them in the same way as for the eutectic system:
$$
g=(C_{\beta}-C_P)/(C_P-C_{\alpha}),\,\,\,
\Delta=(C_{L\beta}^e-C_{L\alpha}^e)/(C_{\beta}-C_{\alpha})
$$
Thus, replacing $-g$ by $|g|$ and $-\Delta$ by $|\Delta|$, we recover  
Eq. (\ref{basic}) with
\begin{equation}\label{rescale1}
X_{\beta}=\frac{x_{\beta}|\Delta|}{\theta_{\beta}^{2}(d_{\beta}+|g|d_{\alpha})},\,\,\,
Y=\frac{y|\Delta|}{\theta_{\beta}(d_{\beta}+|g|d_{\alpha})}
\end{equation}
and 
\begin{equation}\label{parameter1}
\nu=v\frac{|g|(d_{\beta}+|g|d_{\alpha})\theta_{\beta}^{3}}{2D\Delta^2},\,\,\,
\theta_{\beta}=\frac{\varphi_{\alpha}+\varphi_{\beta}}{1-|g|}.
\end{equation} 
Finally, for the steady-state velocity we obtain
\begin{equation}\label{velocity1}
 v=\nu^{\star}\frac{2D\Delta^2}{|g|(d_{\beta}+|g|d_{\alpha})\theta_{\beta}^{3}}
\end{equation}
As we have already noted in the Introduction,   
the interface $x_{\alpha}=|g|x_{\beta}$ represents 
now a solidification front. During the melting, in the peritectic system, 
the low-temperature $\beta$-phase melts while the high-temperature 
$\alpha$-phase solidifies. In addition to the formation of the liquid phase 
the polymorphic transition $\beta\rightarrow\alpha$ takes place. 

\section{Evolution of the solid-solid interface}

We have already mentioned that  diffusion  along the solid-solid interface plays an 
important role allowing the necessary rotation of the structure in the vicinity of 
the triple junction, $x_{\alpha\beta}'(0)=-\theta$ (see Fig. 1b and Fig. 2b). 
At the same time, far away from the triple junction, the solid-solid interface is 
parallel to the $y$ axis ($x_{\alpha\beta}'(\infty)=0$). 
  
The corresponding surface diffusion equation which describes the $x_{\alpha\beta}(y)$
interface has been discussed in \cite{brener03}
and, essentially, it is given by the classical Mullins equation \cite{mul57}:
\begin{equation}\label{mul}
2\delta_{\alpha\beta}d_{\alpha\beta}D_{\alpha\beta}
\frac{d^4x_{\alpha\beta}}{dy^4}-v\frac{dx_{\alpha\beta}}{dy}=0
\end{equation}
In this description the diffusion takes place inside of layer of the thickness 
$2\delta_{\alpha\beta}$ (diffusion coefficient is $D_{\alpha\beta}$) and 
$d_{\alpha\beta}$ is the corresponding capillary length:
$$d_{\alpha\beta}=
\frac{\sigma_{\alpha\beta}\Omega[f_{\alpha}''(C_{\alpha})+ f_{\beta}''(C_{\beta})]}
{2(C_{\beta}-C_{\alpha})^2f_{\alpha}''(C_{\alpha})f_{\beta}''(C_{\beta})}.$$
Here $f_{\alpha(\beta)}(C)$ is the free energy of the $\alpha(\beta)$ phase per atom.
The solution of this equation which has a proper behavior at large values of $y$ 
and satisfies the boundary conditions $x_{\alpha\beta}(0)=0$, 
$x_{\alpha\beta}^{'}(0)=-\theta$ and 
$x_{\alpha\beta}^{''}(0)=\kappa(0)$, is
\begin{eqnarray*}
x_{\alpha\beta}(y) =A-\exp\left(-\frac{qy}{2}\right)
[A\cos (\frac{qy\sqrt{3}}{2})+
B\sin (\frac{qy\sqrt{3}}{2})],
\end{eqnarray*}
\begin{eqnarray}\label{AB}
A=[\kappa(0)/q-\theta]/q;\,\,\,\,B=[\kappa(0)/q+\theta]/(q\sqrt{3}),
\end{eqnarray}
where $q^3=v/(2\delta_{\alpha\beta}d_{\alpha\beta}D_{\alpha\beta})$.
This solution is uniquely defined because the parameters $v$, $\theta$ and $\kappa(0)$
are known from the solution given in the previous sections. The velocity is 
given by Eqs. (\ref{velocity}) and (\ref{velocity1}) for the eutectic and peritectic 
systems respectively; the angle $\theta$ is given by Eq. (\ref{angle}) for both cases   
 and the curvature of the $\alpha/\beta$ interface at the triple junction,
$\kappa(0)\sim \Delta/d_{\alpha\beta}$,  
is also known  since we have already found the 
concentrations at the triple junction. 
The procedure of calculation of $\kappa(0)$ is straightforward but tedious 
and involves new thermodynamical parameters.   
Therefore,  we explain this procedure only schematically:    
The concentrations in the solid phases at the triple junction 
can be calculated, on one hand, using the condition of local equilibrium 
with the liquid phase 
(along the $x_{\alpha(\beta)}$ interface) and, on the other hand, 
using  the condition of local equilibrium between two solid phases (along the 
$x_{\alpha\beta}$ interface). 
At the triple junction these compositions should coincide. The capillary corrections 
to the equilibrium solid concentrations can be written similar to Eq. (\ref{equil})
for all there interfaces. Since the profiles $x_{\alpha}$ and $x_{\beta}$ have been 
already found,  the mentioned continuity condition allows to calculate the curvature
$\kappa(0)$ of the solid-solid interface. 

We note, however, that the term $\kappa(0)/q$ in Eq. (\ref{AB}) is of  order 
$\theta (D_{\alpha\beta}\Delta /D)^{1/3}$ and can be neglected compared to 
$\theta$ for small $\Delta$. 
This means that the tedious calculations explained above are not needed in the limit 
of small $\Delta$. We also note that the characteristic length scales of 
the solid-solid interfacial pattern are small compared to the characteristic 
length scales of the melt structures by the same small parameter 
$(D_{\alpha\beta}\Delta /D)^{1/3}$. Moreover,  the diffusional flux along the 
solid-solid interface has a nonzero value
at the triple junction. In principal,   this flux has to be taken into 
account  in the description of the diffusional field in the liquid phase. This effect 
has been neglected in the previous sections. More careful analysis shows that 
corrections to Eq. (6) due to this effects  are small by the same small parameter, 
 $(D_{\alpha\beta}\Delta /D)^{1/3}<<1.$ Thus, in the limit of small $\Delta$ 
the surface diffusion process  has no influence on the kinetics of the 
contact melting and 
allows only for the necessarily adjustment of the solid-solid interface inside  the 
relatively small region in the vicinity of the triple junction. 

\section{Conclusion}
       
We have developed and analyzed a relatively simple model for the melting kinetics 
along the solid-solid inter-phase  boundaries in eutectic and peritectic systems. 
The process is controlled by the diffusion inside the liquid phase and the existence 
of the triple junction is crucial  for the velocity selection problem.
The additional assumption  of small  opening angles $\varphi$ at the triple junction  
plays only  a technical role and allows to solve the posed problem 
essentially analytically 
using the lubrication approximation for the diffusion field in the liquid phase. 
The obtained scaling results for the dependence of the steady-state velocity of 
the moving pattern on the overheating, which are exact in the limit of small angles,
can still be used  for the moderate values of the angles with only  prefactors 
of the order of unity missing. We note  the obtained scaling relation, 
$v\sim D\Delta^2/d$,  is similar to the scaling relation in eutectic growth.   
Using Eq. (\ref{velocity}) and characteristic values of the parameters 
$D\sim 10^{-9} m^2/s$, $d\sim 10^{-9} m$,   $\Delta\sim 10^{-3}$ and 
$\theta\sim 1$ we estimate the characteristic velocity to be  of the order of 
$v\sim 10^{-6}m/s$. We hope that our results will stimulate some new model 
experiments on contact melting phenomena.

We acknowledge the support by the Deutsche Forschungsgemeinschaft under 
project SPP 1120.

\end{document}